\documentclass[conference]{IEEEtran}
\IEEEoverridecommandlockouts
\usepackage{cite}
\usepackage{amsmath,amssymb,amsfonts}
\usepackage{algorithmic}
\usepackage{graphicx}
\usepackage{textcomp}
\usepackage{xcolor}
\usepackage{booktabs} 
\usepackage{hyperref} 
\def\BibTeX{{\rm B\kern-.05em{\sc i\kern-.025em b}\kern-.08em
    T\kern-.1667em\lower.7ex\hbox{E}\kern-.125emX}}
\begin{document}

\title{Quantum Kernel-Based Long Short-term Memory\\ for Climate Time-Series Forecasting\\

\thanks{\IEEEauthorrefmark{1} Corresponding Author: nanyou@nchc.narl.org.tw (NCHC, Narlabs), \\ \quad kuan-cheng.chen17@imperial.ac.uk (QuEST, Imperial College London)}
}

\author{
\IEEEauthorblockN{
    Yu-Chao Hsu\IEEEauthorrefmark{2}\IEEEauthorrefmark{3},
    Nan-Yow Chen\IEEEauthorrefmark{2}\IEEEauthorrefmark{1},
    Tai-Yu Li\IEEEauthorrefmark{4}, 
    Po-Heng (Henry) Lee \IEEEauthorrefmark{5}\IEEEauthorrefmark{7},
    Kuan-Cheng Chen\IEEEauthorrefmark{5}\IEEEauthorrefmark{6}\IEEEauthorrefmark{1}
}
\IEEEauthorblockA{\IEEEauthorrefmark{2} National Center for High-Performance Computing, NARlabs, Hsinchu, Taiwan}
\IEEEauthorblockA{\IEEEauthorrefmark{3} Cross College Elite Program, National Cheng Kung University, Tainan, Taiwan}
\IEEEauthorblockA{\IEEEauthorrefmark{4}National Synchrotron Radiation Research Center, Hsinchu, Taiwan}
\IEEEauthorblockA{\IEEEauthorrefmark{7}Department of Civil Engineering, Imperial College London, London, UK}
\IEEEauthorblockA{\IEEEauthorrefmark{5}Department of Electrical and Electronic Engineering, Imperial College London, London, UK}
\IEEEauthorblockA{\IEEEauthorrefmark{6}Centre for Quantum Engineering, Science and Technology (QuEST), Imperial College London, London, UK}

}


\maketitle

\begin{abstract}
We present the Quantum Kernel-Based Long Short-Term Memory (QK-LSTM) network, which integrates quantum kernel methods into classical LSTM architectures to enhance predictive accuracy and computational efficiency in climate time-series forecasting tasks, such as Air Quality Index (AQI) prediction. By embedding classical inputs into high-dimensional quantum feature spaces, QK-LSTM captures intricate nonlinear dependencies and temporal dynamics with fewer trainable parameters. Leveraging quantum kernel methods allows for efficient computation of inner products in quantum spaces, addressing the computational challenges faced by classical models and variational quantum circuit-based models. Designed for the Noisy Intermediate-Scale Quantum (NISQ) era, QK-LSTM supports scalable hybrid quantum-classical implementations. Experimental results demonstrate that QK-LSTM outperforms classical LSTM networks in AQI forecasting, showcasing its potential for environmental monitoring and resource-constrained scenarios, while highlighting the broader applicability of quantum-enhanced machine learning frameworks in tackling large-scale, high-dimensional climate datasets.
\end{abstract}


\section{Introduction}
Climate time-series forecasting is essential for understanding and predicting environmental phenomena, which has significant implications for public health\cite{forster2020current}, resource management\cite{haddeland2014global}, and policy-making \cite{ren2023impact}. Accurate forecasting of climatic variables such as temperature, precipitation, and pollutant concentrations enables proactive measures to mitigate adverse effects associated with climate variability and change. Time series forecasting, a fundamental aspect of sequence modeling tasks, is crucial for capturing the temporal dynamics inherent in climate data. Recurrent Neural Networks (RNNs) \cite{RNN} and Long Short-Term Memory (LSTM) networks \cite{LSTM} have been instrumental in addressing these tasks due to their ability to model temporal dependencies within sequential data. However, as the complexity and dimensionality of climate datasets increase—owing to factors such as diverse environmental variables, varying meteorological conditions, and spatial heterogeneity—classical RNNs and LSTMs often require substantial computational resources and extensive parameterization to effectively model intricate patterns and long-range dependencies \cite{yu2019review}.

Quantum computing has emerged as a promising paradigm that leverages quantum mechanical principles such as superposition and entanglement to enhance machine learning models, offering significant computational advantages over traditional methods \cite{huang2021power}. Specifically, quantum machine learning (QML) aims to exploit the computational strengths of quantum systems to process information in high-dimensional Hilbert spaces more efficiently than classical counterparts \cite{peters2021machine,biamonte2017quantum,yu2024shedding,chen2024compressedmediq}. This capability positions quantum computing favorably for large-scale and high-dimensional applications, including environmental monitoring, climate modeling, and other climate-related time-series forecasting tasks \cite{ho2024quantum,nammouchi2023quantum}.

In the context of time series prediction, prior efforts to integrate quantum computing into sequence modeling have led to the development of Quantum-Enhanced Long Short-Term Memory (QLSTM) \cite{chen2022quantum} and Quantum-Trained LSTM \cite{lin2024quantum} architectures based on Variational Quantum Circuits (VQCs) \cite{chen2020variational}. While VQC-based QLSTMs incorporate quantum circuits into neural network structures, they often entail complex circuit designs and necessitate significant quantum resources. This complexity presents substantial challenges for implementation on current quantum hardware, which is constrained by limitations in qubit coherence times and gate fidelities \cite{preskill2018quantum}.

Quantum kernel methods offer an alternative approach by embedding classical data into quantum feature spaces using quantum circuits \cite{blank2020quantum}, enabling efficient computation of inner products (kernels) in these high-dimensional spaces \cite{rebentrost2014quantum,li2015experimental}. Quantum kernels can capture complex data structures with potentially fewer trainable parameters and reduced computational overhead compared to both classical models and VQC-based quantum models \cite{maheshwari2021variational}. This approach leverages the efficiency of quantum systems in representing and manipulating high-dimensional data, providing a means to enhance model expressiveness without proportionally increasing computational demands \cite{gentinetta2024complexity}.

This paper introduces the Quantum Kernel Long Short-Term Memory (QK-LSTM) network \cite{hsu2024quantum}, which integrates quantum kernel methods within the LSTM architecture to improve the modeling of complex sequential patterns in time-series data. By replacing classical linear transformations in the LSTM cells with quantum kernel evaluations, the QK-LSTM leverages quantum feature spaces to more effectively encode intricate dependencies. This integration enables the model to capture non-linear relationships and temporal dynamics that are challenging for classical models to represent, particularly in applications such as Air Quality Index (AQI) prediction, where environmental variables are influenced by a multitude of interdependent factors.

The proposed approach utilizes quantum gates and circuits to perform transformations that are computationally intensive in classical settings, thereby enhancing the efficiency and predictive accuracy of the network for climate forecasting tasks. Furthermore, this integration simplifies the quantum circuit requirements compared to VQC-based QLSTMs, making the QK-LSTM more feasible for implementation on near-term quantum devices and suitable for deployment in quantum edge computing \cite{ma2022hybrid} and resource-constrained environments\cite{fellous2021limitations}. Additionally, the quantum kernel can serve as an effective ansatz for distributed quantum computing, suggesting that this method can be extended towards quantum high-performance computing (HPC) and distributed quantum computing architectures \cite{chen2024consensus,chen2024cutn,burt2024generalised}.

By applying the QK-LSTM to AQI prediction as a case study, we demonstrate its superior performance in capturing the complex temporal dependencies and non-linear patterns inherent in climate time-series data. Experimental results indicate that the QK-LSTM outperforms classical LSTM models in predictive accuracy while requiring fewer parameters. This performance gain highlights the potential of quantum-enhanced models in environmental applications and underscores the viability of integrating quantum computing techniques into existing machine learning frameworks to address computational challenges in processing large-scale, high-dimensional climate data.

\section{Data Pre-processing}

Effective data preprocessing is crucial for improving the performance of time-series models, like those used for predicting AQI\cite{kang2018air}. This involves ensuring data consistency, addressing missing values, and appropriately scaling features. By following these steps, the model can better learn underlying patterns and generate more reliable predictions.

\subsection{Feature Selection}

The AQI is calculated based on various pollutant concentrations \cite{cheng2007comparison}, including carbon monoxide (CO), ammonia (NH$_3$), nitric oxide (NO), nitrogen dioxide (NO$_2$), nitrogen oxides (NO$_x$), sulfur dioxide (SO$_2$), particulate matter with diameters less than or equal to 2.5 micrometers (PM$_{2.5}$) and 10 micrometers (PM$_{10}$), and ozone (O$_3$). These pollutants are critical as they collectively determine the overall air quality, providing a comprehensive representation of environmental health status. Accurate monitoring and analysis of these pollutants are essential for predicting air quality trends and understanding their effects on public health and the environment.

In this study, we focus on Bengaluru, a major city in India. The dataset\cite{cpcb_aqi}, obtained from the Central Pollution Control Board (CPCB) of India, contains several missing values and noisy information. The Xylene feature was excluded from the dataset due to its high rate of missing data and potential to introduce bias. Consequently, a total of 11 features were selected for analysis.

The AQI is computed based on the maximum of the individual pollutant sub-indices, which are normalized concentration values. The AQI calculation is expressed as:

\begin{equation}
    \text{AQI} = \max\left( I_{\text{PM}_{2.5}}, I_{\text{PM}_{10}}, I_{\text{NO}}, I_{\text{NO}_2}, \ldots, I_{\text{NH}_3} \right)
    \label{eq:aqi}
\end{equation}

where \( I_i \) represents the sub-index for pollutant \( i \), obtained by mapping the pollutant's actual concentration \( X_i \) to a standardized scale using established AQI breakpoints. This approach ensures that the AQI reflects the most critical pollutant concentration among the selected features, providing a comprehensive measure of air quality in Bengaluru.

\subsection{Outlier Detection and Removal}

Outlier detection and removal are crucial steps in improving data quality by eliminating anomalous data points that can adversely affect model training. In this study, we employed the Z-score method to detect outliers, which measures how many standard deviations a data point \( x_i \) is from the mean \( \mu \). A data point is considered an outlier if its absolute Z-score exceeds a predefined threshold \( \gamma \):

\begin{equation}
    |Z_i| = \left| \frac{x_i - \mu}{\sigma} \right| > \gamma
    \label{eq:zscore}
\end{equation}

where \( \sigma \) is the standard deviation of the dataset. We set the threshold \( \gamma = 3 \), which corresponds to data points beyond three standard deviations from the mean, a common practice for outlier detection. By removing these outliers, we ensure a more consistent and reliable dataset for modeling.

\subsection{Handling Missing Values}

Due to the extensive number of missing values in the dataset for Bengaluru city, appropriate imputation methods are necessary to maintain data integrity. In this research, we utilized linear interpolation to estimate missing values. Linear interpolation \cite{wong2004comparison} estimates missing data points based on linear relationships between adjacent known data points, which is particularly effective for time-series data. This method minimizes abrupt changes introduced by interpolation, preserving the smoothness and continuity of the dataset.

The linear interpolation formula is expressed as:

\begin{equation}
    f(x) = f(x_0) + \left( \frac{f(x_1) - f(x_0)}{x_1 - x_0} \right) (x - x_0)
    \label{eq:linear_interp}
\end{equation}

where \( x_0 \) and \( x_1 \) are the time points of the known data preceding and succeeding the missing value at time \( x \), and \( f(x_0) \) and \( f(x_1) \) are the corresponding pollutant concentration values. The missing value \( f(x) \) is estimated based on the linear relationship between these points, as shown in \eqref{eq:linear_interp}.

We chose linear interpolation over other imputation methods, such as mean substitution or advanced techniques like spline interpolation, due to its simplicity and effectiveness in handling missing data in time-series without introducing significant bias or complexity.

\section{Methodology}

\subsection{Long Short-Term Memory Networks}

LSTM networks \cite{LSTM} are a specialized form of RNNs \cite{RNN}, designed to address the vanishing and exploding gradient problems commonly encountered in standard RNNs. The unique memory cell structure of LSTMs, comprising input, forget, and output gates, enables the effective retention and management of long-term dependencies in sequential data. LSTMs have been widely adopted in various domains, including natural language processing \cite{chen2016enhanced,yao2018improved,alkin2024vision} and time series forecasting \cite{kilinc2024multimodal,zhou2024self,gulmez2023stock}, due to their ability to capture both short- and long-term relationships within sequences. This capability significantly enhances model performance in sequence-related tasks, particularly in handling long-range dependencies.

\begin{figure}[!b]
    \centering
    \includegraphics[width=0.48\textwidth]{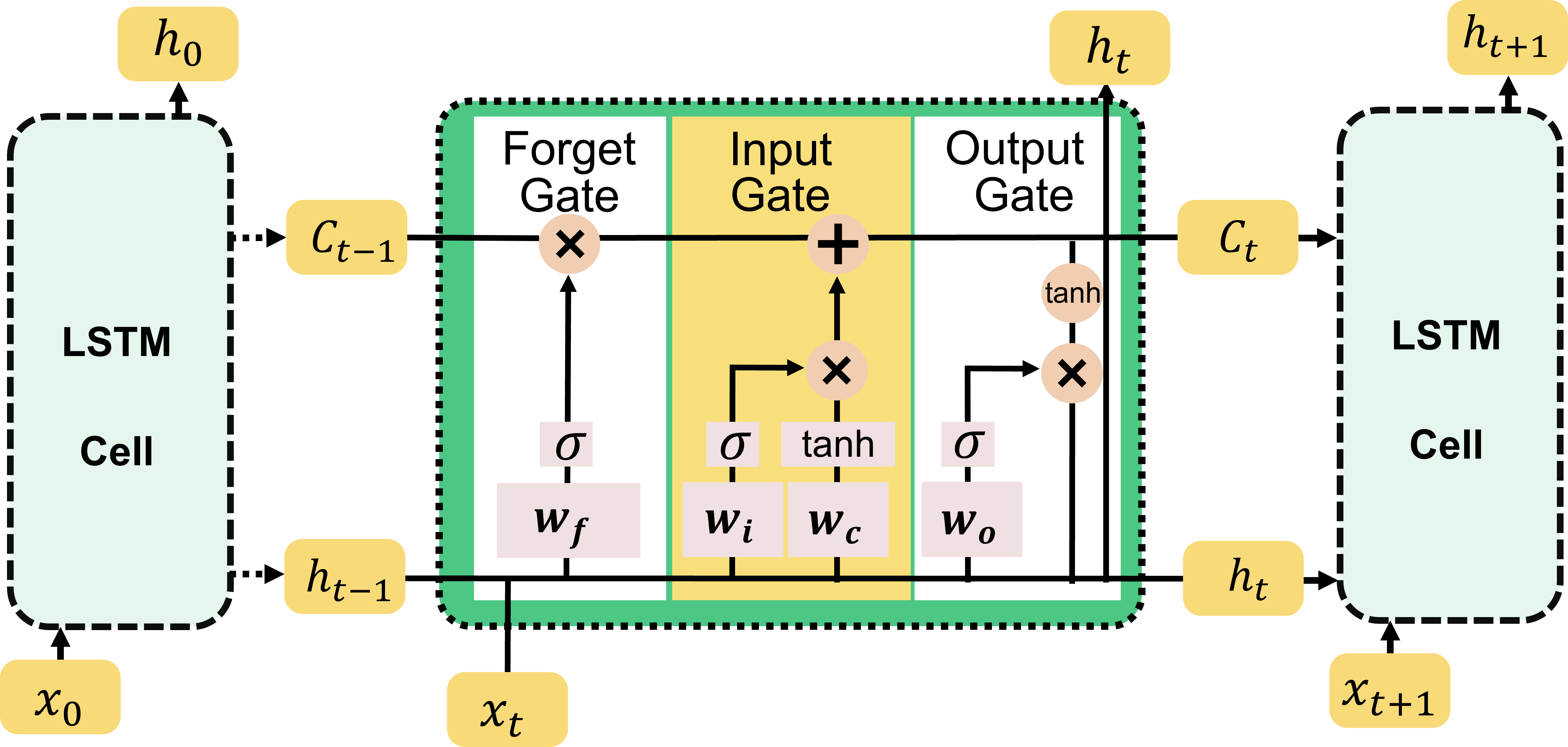}
    \caption{Schematic representation of a standard classical LSTM network architecture.}
    \label{LSTM}
\end{figure}

In this study, we utilize LSTM networks for predicting the AQI, leveraging their proficiency in capturing complex temporal dependencies inherent in climate time-series data. The dynamic and nonlinear characteristics of air quality data, influenced by a multitude of atmospheric and anthropogenic factors, present significant challenges for traditional forecasting methods, which often struggle to provide accurate predictions. The gating mechanisms within LSTM networks offer an effective approach to modeling these complex temporal patterns, enabling the network to capture both seasonal trends and abrupt changes in air quality. This leads to substantial improvements in predictive performance. A schematic representation of a standard classical LSTM network architecture is illustrated in Fig.~\ref{LSTM}.

\subsection{Quantum Kernel-Based Long Short-Term Memory}

\begin{figure*}[!t]
    \centering
    \includegraphics[width=\textwidth]{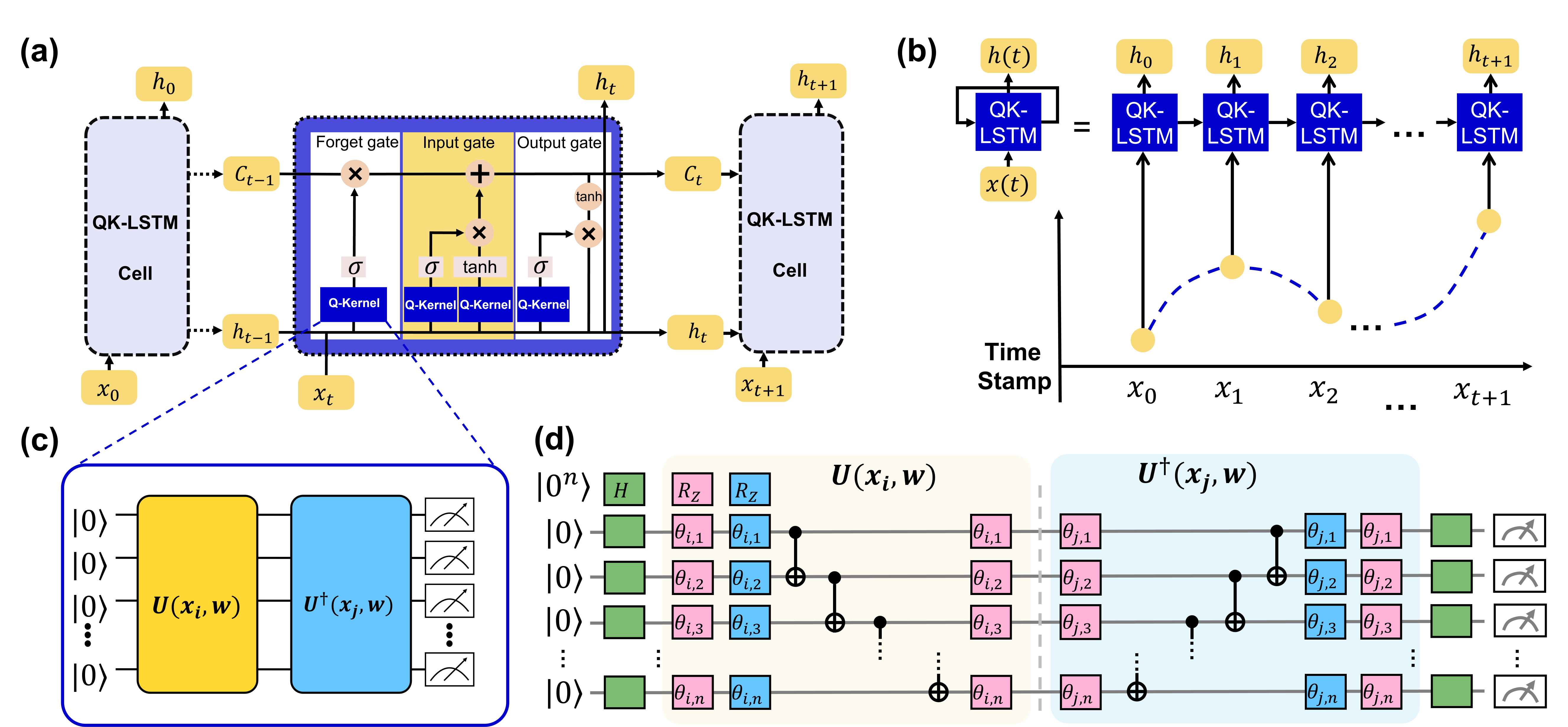}
    \caption{Overview of the QK-LSTM architecture. (a) The QK-LSTM cell combines quantum kernel computations with the traditional LSTM framework, leveraging quantum-enhanced gates (forget, input, and output) to improve sequential data processing and maintain long-term temporal dependencies. (b) The QK-LSTM's sequential structure, where the cell processes input sequences \(x_t\) over multiple time steps, outputs hidden states \(h_t\), and maintains cell states \(C_t\), enabling robust temporal modeling. (c) The unitary operator \(U(x_t, w)\), representing the quantum kernel, encodes classical input data \(x_t\) into a quantum feature space, while its adjoint \(U^{\dagger}(x_j, w)\) computes quantum state overlaps to measure similarity. (d) The quantum circuit representation of \(U(x_t, w)\) and \(U^{\dagger}(x_j, w)\), illustrating the application of quantum gates to encode data and extract features for machine learning tasks. (e) The detailed implementation of the quantum kernel function within the QK-LSTM, showcasing the quantum transformations applied to integrate quantum-enhanced machine learning capabilities.}
    \label{QK-LSTM}
\end{figure*}

To enhance the modeling capability of LSTM networks in capturing intricate nonlinear patterns within sequential climate data, we propose the QK-LSTM model. This model integrates quantum kernel operations within the classical LSTM framework, effectively embedding input data into high-dimensional quantum feature spaces.

As illustrated in Fig.~\ref{QK-LSTM}, the fundamental unit of the proposed QK-LSTM architecture is the QK-LSTM cell. Each QK-LSTM cell modifies the standard LSTM cell by replacing the traditional linear transformations with quantum kernel evaluations. This integration leverages the expressive power of quantum feature spaces to model complex, non-linear relationships in the data, potentially leading to improved predictive performance in climate time-series forecasting tasks such as AQI prediction.

\subsubsection{Classical LSTM Equations}

The standard LSTM cell comprises three gates—the forget gate \( f_t \), the input gate \( i_t \), and the output gate \( o_t \)—and a cell state \( C_t \). The classical LSTM equations governing these components are:
\begin{subequations}
\begin{align}
f_t &= \sigma\left( W_f [h_{t-1}, x_t] + b_f \right), \label{eq:classical_forget_gate} \\
i_t &= \sigma\left( W_i [h_{t-1}, x_t] + b_i \right), \label{eq:classical_input_gate} \\
\tilde{C}_t &= \tanh\left( W_C [h_{t-1}, x_t] + b_C \right), \label{eq:classical_candidate_cell_state} \\
C_t &= f_t \odot C_{t-1} + i_t \odot \tilde{C}_t, \label{eq:classical_cell_state_update} \\
o_t &= \sigma\left( W_o [h_{t-1}, x_t] + b_o \right), \label{eq:classical_output_gate} \\
h_t &= o_t \odot \tanh\left( C_t \right), \label{eq:classical_hidden_state}
\end{align}
\end{subequations}
where:
\begin{itemize}
    \item \( x_t \in \mathbb{R}^n \) is the input vector at time \( t \),
    \item \( h_{t-1} \in \mathbb{R}^m \) is the hidden state from the previous time step,
    \item \( W_f, W_i, W_C, W_o \) are weight matrices,
    \item \( b_f, b_i, b_C, b_o \) are bias vectors,
    \item \( \sigma(\cdot) \) denotes the sigmoid activation function,
    \item \( \tanh(\cdot) \) denotes the hyperbolic tangent activation function,
    \item \( \odot \) represents element-wise multiplication.
\end{itemize}

\subsubsection{Integration of Quantum Kernels into LSTM}

In the QK-LSTM architecture\cite{hsu2024quantum}, we replace the linear transformations \( W [h_{t-1}, x_t] + b \) in the gate computations with quantum kernel evaluations. This approach aims to exploit the high-dimensional representation capabilities of quantum feature spaces to capture complex relationships within the data.

We define the concatenated input vector:
\begin{equation}
v_t = [h_{t-1}; x_t] \in \mathbb{R}^{n + m},
\end{equation}
where \( [\cdot; \cdot] \) denotes vector concatenation.

We introduce a set of reference vectors \( \{ v_j \}_{j=1}^{N} \), which can be a subset of training data or learned during training. The gate activations are computed using weighted sums of quantum kernel functions as follows:

\begin{subequations}
\begin{align}
f_t &= \sigma\left( \sum_{j=1}^{N} \alpha_j^{(f)} k^{(f)}(v_t, v_j) + b_f \right), \label{eq:qk_forget_gate} \\
i_t &= \sigma\left( \sum_{j=1}^{N} \alpha_j^{(i)} k^{(i)}(v_t, v_j) + b_i \right), \label{eq:qk_input_gate} \\
\tilde{C}_t &= \tanh\left( \sum_{j=1}^{N} \alpha_j^{(C)} k^{(C)}(v_t, v_j) + b_C \right), \label{eq:qk_candidate_cell_state} \\
C_t &= f_t \odot C_{t-1} + i_t \odot \tilde{C}_t, \label{eq:qk_cell_state_update} \\
o_t &= \sigma\left( \sum_{j=1}^{N} \alpha_j^{(o)} k^{(o)}(v_t, v_j) + b_o \right), \label{eq:qk_output_gate} \\
h_t &= o_t \odot \tanh\left( C_t \right), \label{eq:qk_hidden_state}
\end{align}
\end{subequations}

where:
\begin{itemize}
    \item \( \alpha_j^{(f)} \), \( \alpha_j^{(i)} \), \( \alpha_j^{(C)} \), and \( \alpha_j^{(o)} \) are trainable weights associated with the quantum kernels for each gate,
    \item \( k^{(f)}(\cdot, \cdot) \), \( k^{(i)}(\cdot, \cdot) \), \( k^{(C)}(\cdot, \cdot) \), and \( k^{(o)}(\cdot, \cdot) \) are quantum kernel functions specific to each gate,
    \item \( b_f \), \( b_i \), \( b_C \), and \( b_o \) are bias terms.
\end{itemize}

\subsubsection{Quantum Kernel Function}

The quantum kernel function \( k(v_t, v_j) \) measures the similarity between two data points \( v_t \) and \( v_j \) in a quantum feature space induced by a quantum feature map \( \phi(v) \). It is defined as:
\begin{equation}
k(v_t, v_j) = \left| \langle \phi(v_t) | \phi(v_j) \rangle \right|^2,
\end{equation}
where \( | \phi(v) \rangle \) represents the quantum state corresponding to the classical input \( v \).

The quantum feature map \( \phi(v) \) is implemented via a parameterized quantum circuit \( U(v) \) that encodes the classical data \( v \) into a quantum state:
\begin{equation}
| \phi(v) \rangle = U(v) | 0 \rangle^{\otimes n},
\end{equation}
with \( n \) being the number of qubits.

\paragraph{Quantum Circuit Design}

The quantum circuit \( U(v) \) comprises the following components:

\begin{enumerate}
    \item \textbf{Initialization}: All qubits are initialized to the \( | 0 \rangle \) state.
    \item \textbf{Hadamard Transformation}: Apply Hadamard gates to create a superposition:
    \begin{equation}
    | \psi_0 \rangle = H^{\otimes n} | 0 \rangle^{\otimes n}.
    \end{equation}
    \item \textbf{Data Encoding}: Encode classical data using parameterized rotation gates:
    \begin{equation}
    U_{\text{enc}}(v) = \prod_{k=1}^{n} R_y(\theta_{k}) R_z(\phi_{k}),
    \end{equation}
    where \( \theta_{k} \) and \( \phi_{k} \) are functions of the components of \( v \).
    \item \textbf{Entanglement}: Introduce entanglement using controlled-NOT (CNOT) gates:
    \begin{equation}
    U_{\text{ent}} = \prod_{k=1}^{n-1} \text{CNOT}(k, k+1).
    \end{equation}
    \item \textbf{Final State Preparation}: The quantum state after encoding is:
    \begin{equation}
    | \phi(v) \rangle = U_{\text{ent}} U_{\text{enc}}(v) | \psi_0 \rangle.
    \end{equation}
\end{enumerate}

\paragraph{Quantum Kernel Evaluation}

The quantum kernel between \( v_t \) and \( v_j \) is evaluated as:
\begin{equation}
k(v_t, v_j) = \left| \langle 0 |^{\otimes n} U^\dagger(v_j) U(v_t) | 0 \rangle^{\otimes n} \right|^2.
\end{equation}

This computation involves preparing the quantum states corresponding to \( v_t \) and \( v_j \), applying the inverse circuit \( U^\dagger(v_j) \) followed by \( U(v_t) \), and measuring the probability of the system being in the \( | 0 \rangle^{\otimes n} \) state. This procedure effectively computes the inner product between the quantum states \( | \phi(v_t) \rangle \) and \( | \phi(v_j) \rangle \), capturing their similarity in the quantum feature space.

\subsubsection{Training and Optimization}

The parameters of the QK-LSTM model include the weights \( \alpha_j^{(f)} \), \( \alpha_j^{(i)} \), \( \alpha_j^{(C)} \), \( \alpha_j^{(o)} \), biases \( b_f \), \( b_i \), \( b_C \), \( b_o \), and any trainable parameters within the quantum circuits used for the kernel computations.

\paragraph{Loss Function}

For the regression task of AQI prediction, we define a suitable loss function, such as the Mean Squared Error (MSE):
\begin{equation}
\mathcal{L} = \frac{1}{T} \sum_{t=1}^{T} \left( y_t - \hat{y}_t \right)^2,
\end{equation}
where \( y_t \) is the true AQI value at time \( t \), \( \hat{y}_t \) is the predicted AQI value, and \( T \) is the total number of time steps.

\paragraph{Gradient Computation}

The gradients of the loss with respect to the classical parameters \( \alpha_j \) and biases \( b \) are computed using standard backpropagation through time (BPTT). For the quantum circuit parameters, we employ the parameter-shift rule \cite{parameter-shift}, which allows for efficient computation of gradients in quantum circuits.

\paragraph{Parameter-Shift Rule}

The gradient of the quantum kernel with respect to a circuit parameter \( \theta \) is given by:
\begin{equation}
\frac{\partial k(v_t, v_j)}{\partial \theta} = k_{\theta}^{+}(v_t, v_j) - k_{\theta}^{-}(v_t, v_j),
\end{equation}
where \( k_{\theta}^{\pm}(v_t, v_j) \) represents the kernel evaluated with the parameter \( \theta \) shifted by \( \pm \frac{\pi}{2} \):
\begin{equation}
k_{\theta}^{\pm}(v_t, v_j) = \left| \langle 0 |^{\otimes n} U^\dagger(v_j) U_{\theta \pm \frac{\pi}{2}}(v_t) | 0 \rangle^{\otimes n} \right|^2.
\end{equation}

\paragraph{Optimization Algorithm}

An optimization algorithm, such as stochastic gradient descent (SGD) or Adam, is employed to update the parameters:
\begin{align}
\alpha_j &\leftarrow \alpha_j - \eta \frac{\partial \mathcal{L}}{\partial \alpha_j}, \\
b &\leftarrow b - \eta \frac{\partial \mathcal{L}}{\partial b}, \\
\theta &\leftarrow \theta - \eta \frac{\partial \mathcal{L}}{\partial \theta},
\end{align}
where \( \eta \) is the learning rate.

\section{Result}

\begin{figure*}[t]
    \vspace*{-\intextsep}
    \centering
    \includegraphics[width=1 \textwidth]{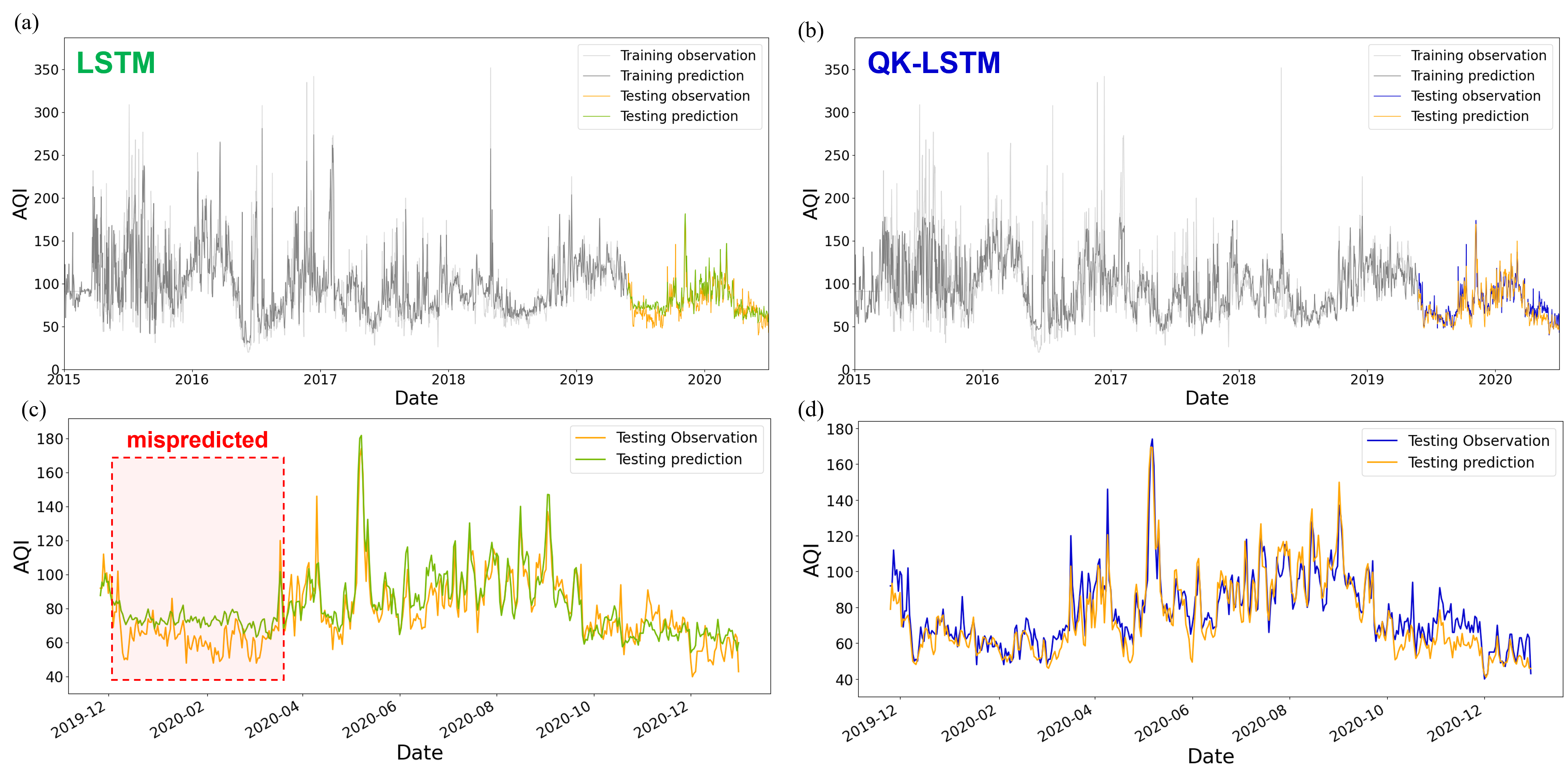}
\caption{Comparative analysis of air quality prediction performance using classical LSTM and QK-LSTM models. Panels (a) and (b) depict the overall predictive capabilities of the classical LSTM and QK-LSTM models, respectively, across the full dataset (2015-2020), including both training and testing phases. Panels (c) and (d) focus on the models' testing performance for 2020, with (c) highlighting the classical LSTM and (d) showcasing the QK-LSTM. The results demonstrate the models' ability to capture seasonal variations and rapid changes in air quality, with QK-LSTM exhibiting enhanced fidelity in tracking testing observations.}
    \label{Result}
\end{figure*}

\subsection{Evaluation Method}

To rigorously assess the performance of the proposed QK-LSTM model, we employ a suite of evaluation metrics that provide a comprehensive and quantitative analysis of the regression model's accuracy and error characteristics. These metrics facilitate an objective comparison between the QK-LSTM and traditional LSTM models.

\subsubsection{Root Mean Square Error (RMSE)}

The Root Mean Square Error (RMSE) is a widely used metric that measures the average magnitude of the prediction errors. It is defined as:

\begin{equation}
    \text{RMSE} = \sqrt{\frac{1}{N} \sum_{i=1}^{N} \left( y_i - \hat{y}_i \right)^2}
    \label{eq:rmse}
\end{equation}

where \( N \) denotes the total number of observations, \( y_i \) represents the actual AQI values, and \( \hat{y}_i \) denotes the predicted AQI values by the model. RMSE provides a measure of the model's prediction accuracy, with lower values indicating superior performance.

\subsubsection{Mean Absolute Error (MAE)}

The Mean Absolute Error (MAE) quantifies the average absolute differences between the predicted and actual values. It is defined as:

\begin{equation}
    \text{MAE} = \frac{1}{N} \sum_{i=1}^{N} \left| y_i - \hat{y}_i \right|
    \label{eq:mae}
\end{equation}

MAE offers a straightforward interpretation of the average prediction error, with lower values indicating higher accuracy. Unlike RMSE, MAE is less sensitive to large deviations, providing a robust measure of model performance.

\subsubsection{Mean Absolute Percentage Error (MAPE)}

The Mean Absolute Percentage Error (MAPE) expresses the prediction error as a percentage of the actual values, enhancing interpretability. It is calculated as:

\begin{equation}
    \text{MAPE} = \frac{100}{N} \sum_{i=1}^{N} \left| \frac{y_i - \hat{y}_i}{y_i} \right|
    \label{eq:mape}
\end{equation}

MAPE facilitates the understanding of the model's error relative to the magnitude of the actual values, making it particularly useful for comparing performance across different scales. Lower MAPE values indicate better predictive performance.

\subsubsection{Coefficient of Determination}

The coefficient of determination (\( R^2 \)) measures the proportion of variance in the dependent variable that is predictable from the independent variables. It is defined as:

\begin{equation}
    R^2 = 1 - \frac{\sum_{i=1}^{N} \left( y_i - \hat{y}_i \right)^2}{\sum_{i=1}^{N} \left( y_i - \bar{y} \right)^2}
    \label{eq:r2}
\end{equation}

where \( \bar{y} \) represents the mean of the actual AQI values. The \( R^2 \) value ranges from 0 to 1, with values closer to 1 indicating that a greater proportion of variance is explained by the model, thereby reflecting better performance.

\subsubsection{Performance Benchmarking}

\begin{table}[!b]
    \caption{Performance Benchmarking Between QK-LSTM and LSTM}
    \label{tab:performance_comparison}
    \centering
    \begin{tabular}{@{}cccc@{}}
        \toprule
        \textbf{Metric} & \textcolor{blue}{\textbf{QK-LSTM}} & \textbf{LSTM} & \textbf{Improvement} \\
        \midrule
        RMSE & \textbf{9.20} & 15.94 & -42.8\% \\
        MAE       & \textbf{7.15} & 11.07 & -35.4\% \\
        MAPE & \textbf{9.14\%} & 13.32\% & -31.9\% \\
        \( R^2 \)           & \textbf{0.84} & 0.78 & +7.1\% \\
        \bottomrule
    \end{tabular}
\end{table}

To provide a clear comparison between the QK-LSTM and the conventional LSTM models, we present the evaluation metrics in Table~\ref{tab:performance_comparison}. The results demonstrate the effectiveness of the QK-LSTM in improving prediction accuracy while maintaining computational efficiency.

\noindent In Table~\ref{tab:performance_comparison}, the QK-LSTM model demonstrates substantial reductions in RMSE, MAE, and MAPE compared to the traditional LSTM model, highlighting its enhanced predictive accuracy. Additionally, the $R^{2}$ value is slightly higher for the QK-LSTM, further reinforcing the model's ability to capture intricate temporal dependencies and non-linear patterns within air quality data.

\subsection{Model Compression}

\begin{table}[!b]
    \caption{Comparison of Parameters for QK-LSTM and LSTM Networks.}
    \label{tab:quantum_lstm_vs_lstm_parameters}
    \centering
    \begin{tabular}{@{}ccc@{}}
        \toprule
        \multicolumn{1}{c}{\textbf{Hyperparameter}} & \textcolor{blue}{\textbf{QK-LSTM}} & \textbf{LSTM} \\
        \midrule
        Epochs & 20 & 20 \\
        Learning Rate & 0.001 & 0.001 \\
        Number of hidden units & 16 & 16 \\
        Batch size & 1 & 1 \\
        Sequence Length & 3 & 3 \\
        Number of Qubits in Quantum Kernel Circuit & 4 & -- \\
        \midrule
        \textbf{Total Trainable Parameters} & \textbf{209} & 1873 \\
        \bottomrule
    \end{tabular}
\end{table}

Table~\ref{tab:quantum_lstm_vs_lstm_parameters} presents a comparative analysis of the hyperparameters and total trainable parameters between the proposed QK-LSTM network and the traditional LSTM network. Both models were configured with identical settings, including 20 epochs, a learning rate of 0.001, 16 hidden units, a batch size of 1, and a sequence length of 3. The primary distinction lies in the integration of a quantum kernel circuit within the QK-LSTM, utilizing 4 qubits, whereas the conventional LSTM does not incorporate any quantum components. Notably, the QK-LSTM model requires only 209 trainable parameters compared to 1,873 parameters in the standard LSTM. This substantial reduction in the number of parameters highlights the efficiency of the QK-LSTM in capturing complex data representations through quantum-enhanced feature spaces. Fewer trainable parameters not only decrease the computational resource requirements and training time but also mitigate the risk of overfitting, thereby enhancing the model's generalization capabilities. The incorporation of quantum circuits enables the QK-LSTM to model high-dimensional dependencies and non-linear patterns more effectively with a streamlined architecture. This efficiency makes the QK-LSTM particularly advantageous for deployment in resource-constrained environments and supports scalability for large-scale applications. Overall, the performance benchmarking underscores the potential of quantum-enhanced neural networks to achieve comparable or superior predictive performance with significantly reduced model complexity.

\section{Scalability and Practicality}

The scalability and practicality of the QK-LSTM model are critical considerations for its application to large-scale, high-dimensional climate time-series forecasting tasks. The theoretical foundations of QK-LSTM lie in the integration of quantum kernel methods with classical LSTM architectures, combining the strengths of quantum computing in handling complex data structures with the temporal modeling capabilities of LSTMs.

The QK-LSTM model has been successfully applied to tasks such as Part-of-Speech (POS) tagging task \cite{hsu2024quantum}, demonstrating that the model achieved competitive accuracy while significantly reducing the number of trainable parameters compared to classical LSTM models. The reduction in parameters is theoretically advantageous, as it decreases the computational complexity and mitigates the risk of overfitting. This efficiency stems from the quantum kernel's ability to implicitly map input data into high-dimensional Hilbert spaces, allowing the model to capture intricate patterns without the need for extensive parameterization \cite{schuld2019quantum}.

To further assess the scalability and applicability of QK-LSTM across domains, we extended its use to Air Quality Index (AQI) prediction—a task characterized by complex temporal dynamics and nonlinear dependencies. The experimental results indicated that QK-LSTM outperforms classical LSTM models in predictive accuracy while requiring significantly fewer trainable parameters. Theoretically, this performance gain is attributed to the quantum kernel's capacity to efficiently compute inner products in high-dimensional feature spaces, effectively enhancing the model's expressiveness \cite{havlivcek2019supervised}. This property is particularly beneficial for modeling climate time-series data, which often exhibit nonlinearity and high dimensionality due to the multitude of influencing factors.

The QK-LSTM model leverages a specially designed quantum circuit employing the block-encoding technique \cite{martyn2020entanglement,suzuki2024quantum}, which facilitates efficient quantum kernel computations. Block-encoding allows for the representation of complex operators within a larger unitary matrix, enabling efficient implementation of quantum kernels that can handle large-scale data. The theoretical advantage of block-encoding lies in its ability to approximate functions of large matrices, such as kernel matrices, without explicitly constructing them, thus reducing computational overhead \cite{gilyen2019quantum}.

Moreover, the QK-LSTM is designed with the Noisy Intermediate-Scale Quantum (NISQ) era in mind. Recognizing the current limitations of quantum hardware, such as qubit coherence times and gate fidelities \cite{preskill2018quantum}, the model allows for parts of the quantum kernel computations to be simulated on classical hardware, particularly GPUs \cite{chen2024cutn}. This approach is theoretically supported by the correspondence between certain quantum computations and tensor network contractions, which can be efficiently executed on GPUs due to their parallel processing capabilities \cite{markov2008simulating}. By simulating quantum kernels on classical hardware, the QK-LSTM effectively balances resource demands and mitigates the limitations of current quantum devices.

\section{Conclusion}
This study has demonstrated that integrating quantum kernel methods into LSTM networks significantly enhances climate time-series forecasting, as exemplified by Air Quality Index (AQI) prediction. The proposed QK-LSTM model leverages high-dimensional quantum feature spaces to capture complex nonlinear temporal dependencies inherent in climate data, resulting in improved predictive accuracy and reduced numbers of trainable parameters, thereby increasing computational efficiency. This approach aligns with advancements in quantum high-performance computing (HPC) and hybrid quantum algorithm frameworks, facilitating scalable implementations on emerging quantum hardware and classical co-processors. The successful application of QK-LSTM to AQI prediction suggests its broader applicability to other climate change and time-series analysis problems, such as temperature forecasting, precipitation prediction, flood modeling, and greenhouse gas emission analysis. By extending the QK-LSTM framework to these domains, researchers can effectively address the challenges posed by the intricate and nonlinear nature of climate data, ultimately contributing to more accurate and reliable climate modeling. This advancement supports enhanced decision-making processes related to environmental management and policy formulation, highlighting the potential of quantum-enhanced machine learning models within the context of quantum HPC and hybrid quantum computing paradigms for tackling large-scale, high-dimensional datasets.



\section*{Acknowledgment}
This work was financially supported by the National Science and Technology Council (NSTC), Taiwan, under Grant NSTC 112-2119-M-007-008- and 113-2119-M-007-013- and EPSRC Distributed Quantum Computing and Applications project (grant number EP/W032643/1). The authors would like to thank the National Center for High-performance Computing of Taiwan for providing computational and storage resources.The authors also thank Dr. An-Cheng Yang and Dr. Chun-Yu Lin for their support with the hardware environment and for the valuable discussions.

\bibliographystyle{ieeetr}
\bibliography{reference} 

\vspace{12pt}

\end{document}